# Title: Respiration driven CO$_2$ pulses dominate Australia's flux variability

**Authors:** Eva-Marie Metz[1*], Sanam N. Vardag[1,2], Sourish Basu[3,4], Martin Jung[5], Bernhard Ahrens[5], Tarek El-Madany[5], Stephen Sitch[6], Vivek K. Arora[7], Peter R. Briggs[8], Pierre Friedlingstein[9,10], Daniel S. Goll[11], Atul K. Jain[12], Etsushi Kato[13], Danica Lombardozzi[14], Julia E.M.S. Nabel[5,15], Benjamin Poulter[16], Roland Séférian[17], Hanqin Tian[18], Andrew Wiltshire[19], Wenping Yuan[20], Xu Yue[21], Sönke Zaehle[5], Nicholas M. Deutscher[22], David W.T. Griffith[22] and André Butz[1,2,23*]

**Affiliations:**

[1]Institute of Environmental Physics, Heidelberg University; Im Neuenheimer Feld 229, Heidelberg, 69120, Germany.

[2]Heidelberg Center for the Environment (HCE), Heidelberg University; Im Neuenheimer Feld, 130.1, Heidelberg, 69120, Germany.

[3]Goddard Space Flight Center, NASA; 8800 Greenbelt Rd, Greenbelt, 20771, Maryland, USA.

[4]Earth System Science Interdisciplinary Center, University of Maryland; 5825 University Research Court, Suite 4001, College Park, 20740, Maryland, USA.

[5]Max Planck Institute for Biogeochemistry; Hans-Knöll-Straße 10, Jena, 07745, Germany.

[6]College of Life and Environmental Sciences, University of Exeter; North Park Road, Exeter, EX4 4RJ, Devon, UK.

[7]Canadian Centre for Climate Modelling and Analysis, Environment and Climate Change Canada; 200-2474 Arbutus Road, Victoria, B.C., V8N 1V9, Canada.

[8]Climate Science Centre, CSIRO Oceans and Atmosphere; Canberra, ACT 2601, Australia.

[9]College of Engineering, Mathematics and Physical Sciences, University of Exeter; Exeter EX4 4QF, United Kingdom.

[10]Laboratoire de Météorologie Dynamique, Institut Pierre-Simon Laplace, CNRS-ENS-UPMC-X, Département de Géosciences, Ecole Normale Supérieure; 24 rue Lhomond, 75005 Paris, France.

[11]Université Paris Saclay, CEA-CNRS-UVSQ, LSCE/IPSL; Gif sur Yvette, France.

[12]Department of Atmospheric Sciences, University of Illinois; Urbana, IL 61801, USA.

[13]Institute of Applied Energy; Tokyo 105-0003, Japan.

[14]Climate and Global Dynamics Laboratory, National Center for Atmospheric Research; 1850 Table Mesa Drive Boulder, CO 80305, USA

[15]Max Planck Institute for Meteorology; Bundesstr. 53, 20146 Hamburg, Germany.

[16]Biospheric Sciences Laboratory, Goddard Space Flight Center, NASA; Greenbelt, 20771, Maryland, USA.

[17]CNRM, Université de Toulouse, Météo-France, CNRS; Toulouse, France.

[18]International Center for Climate and Global Change Research, School of Forestry and Wildlife Sciences, Auburn University; AL 36849, USA.

[19]Met Office Hadley Centre; FitzRoy Road, Exeter EX1 3PB, UK.

[20]School of Atmospheric Sciences, Southern Marine Science and Engineering Guangdong Laboratory (Zhuhai), Sun Yat-sen University; Zhuhai 519082, China.

[21]School of Environmental Science and Engineering, Nanjing University of Information Science & Technology (NUIST); Nanjing 210044, China.

[22]Centre for Atmospheric Chemistry, School of Chemistry, University of Wollongong; Wollongong, NSW, 2522, Australia.

[23]Interdisciplinary Center for Scientific Computing (IWR), Heidelberg University; Im Neuenheimer Feld, 205, Heidelberg, 69120, Germany.

*Corresponding author. Email: eva-marie.metz@iup.uni-heidelberg.de, andre.butz@iup.uni-heidelberg.de

**Abstract:** The Australian continent contributes substantially to the year-to-year variability of the global terrestrial carbon dioxide ($CO_2$) sink. However, the scarcity of in-situ observations in remote areas prevents deciphering the processes that force the $CO_2$ flux variability. Here, examining atmospheric $CO_2$ measurements from satellites in the period 2009-2018, we find recurrent end-of-dry-season $CO_2$ pulses over the Australian continent. These pulses largely control the year-to-year variability of Australia's $CO_2$ balance, due to 2-3 times higher seasonal variations compared to previous top-down inversions and bottom-up estimates. The $CO_2$ pulses occur shortly after the onset of rainfall and are driven by enhanced soil respiration preceding photosynthetic uptake in Australia's semi-arid regions. The suggested continental-scale relevance of soil rewetting processes has large implications for our understanding and modelling of global climate-carbon cycle feedbacks.

**One-Sentence Summary:** Satellite $CO_2$ measurements find large $CO_2$ pulses over Australia attributable to rewetting of seasonally dry soils.

**Main Text:** Terrestrial ecosystems drive the seasonal and year-to-year variability of the global carbon dioxide ($CO_2$) sink (*1*). Previous research identified semi-arid regions as hotspots of global $CO_2$ balance inter-annual variability (*2–5*) due to their large sensitivity of photosynthetic carbon uptake to fluctuations in water availability (*6, 7*). The Australian continent is primarily covered with semi-arid ecosystems and experiences large variations in rainfall. This makes Australia particularly relevant for the variability in the global carbon cycle (*8–13*), contributing up to 60% to yearly anomalies of the global terrestrial $CO_2$ sink (*2*).

However, current approaches for attributing global $CO_2$ sink variations to certain regions and mechanisms are highly uncertain, which limits our ability to model climate-carbon cycle feedbacks and project future climate (*14, 15*). Global process-based ecosystem models underestimate observed $CO_2$ flux variability across semi-arid sites due to the complexity of carbon-water cycle interactions and the diversity of ecosystem responses to water fluctuations (*16, 17*). The same holds true for machine learning based models trained on local carbon flux observations (*18, 19*), which is due to the scarcity of available flux measurements in low-latitude semi-arid regions (*20*) as well as due to the inability to represent potentially important non-instantaneous carry-over effects (*21*). Atmospheric transport inversions based on in-situ measurements of airborne $CO_2$ also suffer from the scarcity of observations in remote areas and thus the inversions cannot reliably attribute $CO_2$ flux variability to specific regions, despite growing monitoring capacities (*22, 23*). However, recent satellite observations of atmospheric column $CO_2$ deliver data where ground-based in-situ concentration measurements and carbon flux networks are sparse and thus, satellite $CO_2$ data can fill important gaps and provide new constraints on regional scale patterns and processes (*8, 24–28*).

Here, using satellite observations of atmospheric $CO_2$ concentrations from the Greenhouse Gases Observing Satellite (GOSAT) for the period 2009 to 2018, we identify a net $CO_2$ pulse to the atmosphere that occurs over Australia at the end of the dry season in most years with variable magnitude. We show that this pattern appears to dominate the seasonal and year-to-year variations of Australia's $CO_2$ balance for that period, while it is not evident in traditional atmospheric inversions using in-situ measurements only, in the FLUXCOM machine learning based extrapolations of in-situ flux measurements, and most process-based ecosystem models of the TRENDY initiative. The few process-based TRENDY models that reproduce the $CO_2$ pulse pattern qualitatively suggest that it is caused by rapid respiratory carbon release with the onset of the rainy season while the increase in photosynthetic carbon uptake lags behind. This observed process is consistent with the phenomenon of respiration pulses after rewetting events known as "Birch effect" (*29, 30*) The Birch effect has been described extensively in local studies of water-limited systems (*31*) but its large-scale relevance remained unknown.

**Atmospheric $CO_2$ peak over Australia**

The Greenhouse Gases Observing Satellite (GOSAT) has been delivering global measurements of the column-average dry-air mole fractions ("concentrations") of atmospheric $CO_2$ since its launch in 2009 (*32*). After subtracting the secular trend (*33*), the record of GOSAT record for the period 2009-2018 (Fig. 1) reveals a seasonal pattern above Australia with $CO_2$ draw-down in March, April, May (MAM) and a $CO_2$ peak of variable magnitude at the end of the dry season in October, November, December (OND). These patterns are consistent among two retrievals independently applied to GOSAT (GOSAT/RemoTeC (*34*) and GOSAT/ACOS (*35*), Table S1) and they are present in $CO_2$ concentrations measured by the Orbiting Carbon Observatory (OCO-2 (*36, 37*), period 2014 to 2018, Table S1) as well as in ground-based data of the Total Carbon Column Observing Network (*38*) (Fig. S1 and S2).

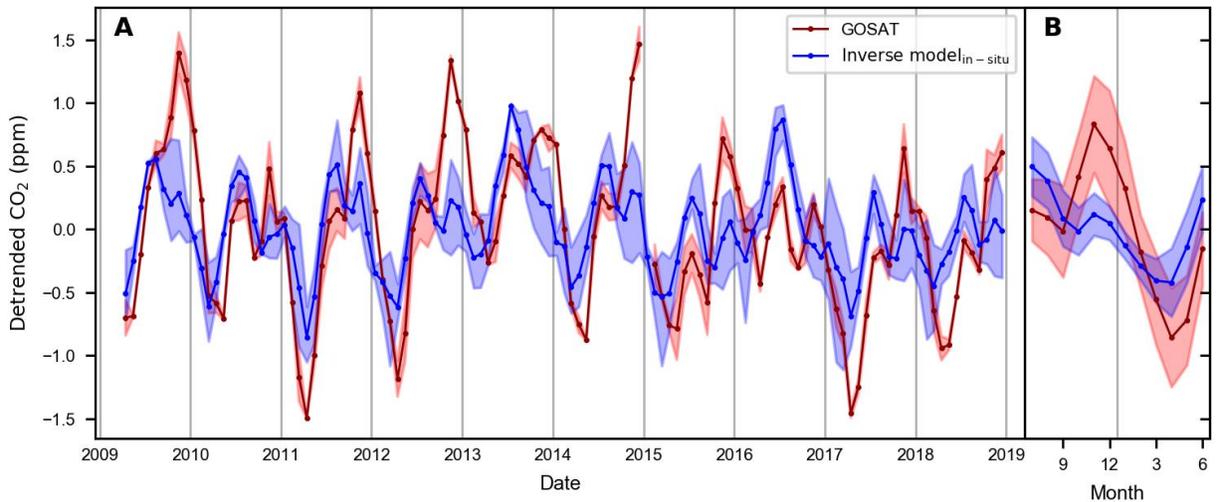

**Fig. 1. Detrended CO$_2$ concentrations over Australia from satellite and models. (A)** Detrended column-average dry-air mole fractions of CO$_2$ measured by GOSAT (red) and simulated by inverse models assimilating in-situ ground-based measurements (blue). Data are monthly averages for Australia. Red shading indicates the range of the GOSAT/RemoTeC and GOSAT/ACOS algorithms. Blue shading indicates the range of the CarbonTracker, CAMS, and TM5-4DVAR inverse models. **(B)** Mean and standard deviation (shading) over the period 2009 to 2018.

In contrast, the atmospheric column CO$_2$ concentrations simulated by three inverse atmospheric transport models (CarbonTracker CT2019B (*39*), CAMS (*40*), TM5-4DVAR (*41*)) underestimate the CO$_2$ draw-down in MAM and lack the CO$_2$ pulses in OND (Fig. 1). Driven by atmospheric winds, these transport models deliver concentration fields that are optimally compatible with in-situ measured CO$_2$ concentrations and the a priori biogenic, oceanic, fire and fossil CO$_2$ surface-atmosphere fluxes (*33*). However, due to their sparsity in and around Australia (see Fig. S3), the in-situ measurements provide only marginal constraints on the regional flux balance. Thus, the discrepancy between CO$_2$ concentrations from GOSAT and traditional in-situ based atmospheric inversions hints at the existence of a carbon release mechanism in Australian ecosystems that has remained undetected by the existing in-situ $CO_2$ monitoring system.

**Australian top-down and bottom-up fluxes**

To improve on the surface flux estimates for Australia, we feed the GOSAT $CO_2$ concentrations into one of the atmospheric inverse models (TM5-4DVAR) together with the in-situ $CO_2$ measurements. We find indeed that the recurring end-of-dry-season $CO_2$ concentration peaks are attributed to a carbon release pattern originating from land ecosystems, which is not present in the inversions when assimilating in-situ $CO_2$ data alone (Fig. 2A).

Our new estimates of Australia's carbon balance variability based on assimilating GOSAT together with in-situ data show a nearly doubled peak-to-peak amplitude of the seasonal cycle (172±47 TgC/month, mean ± standard deviation over the 2009 to 2018 period, July-to-June peak-to-peak amplitude) compared to the in-situ-only inversions (93±11 TgC/month). Moreover, the end-of-dry-season $CO_2$ pulses found by the GOSAT inversions imply a 4-fold greater year-to-year variability of the annual $CO_2$ fluxes (0.233 PgC/a, standard deviation over the 2010 to 2018 period) than for the in-situ-only inversions (0.048 PgC/a) (Table S2). Fluxes obtained by assimilating OCO-2 together with in-situ data for the period 2015 to 2018 show the same end-of-dry-season pulses and agree well with the fluxes of the GOSAT inversion (see Fig. S4).

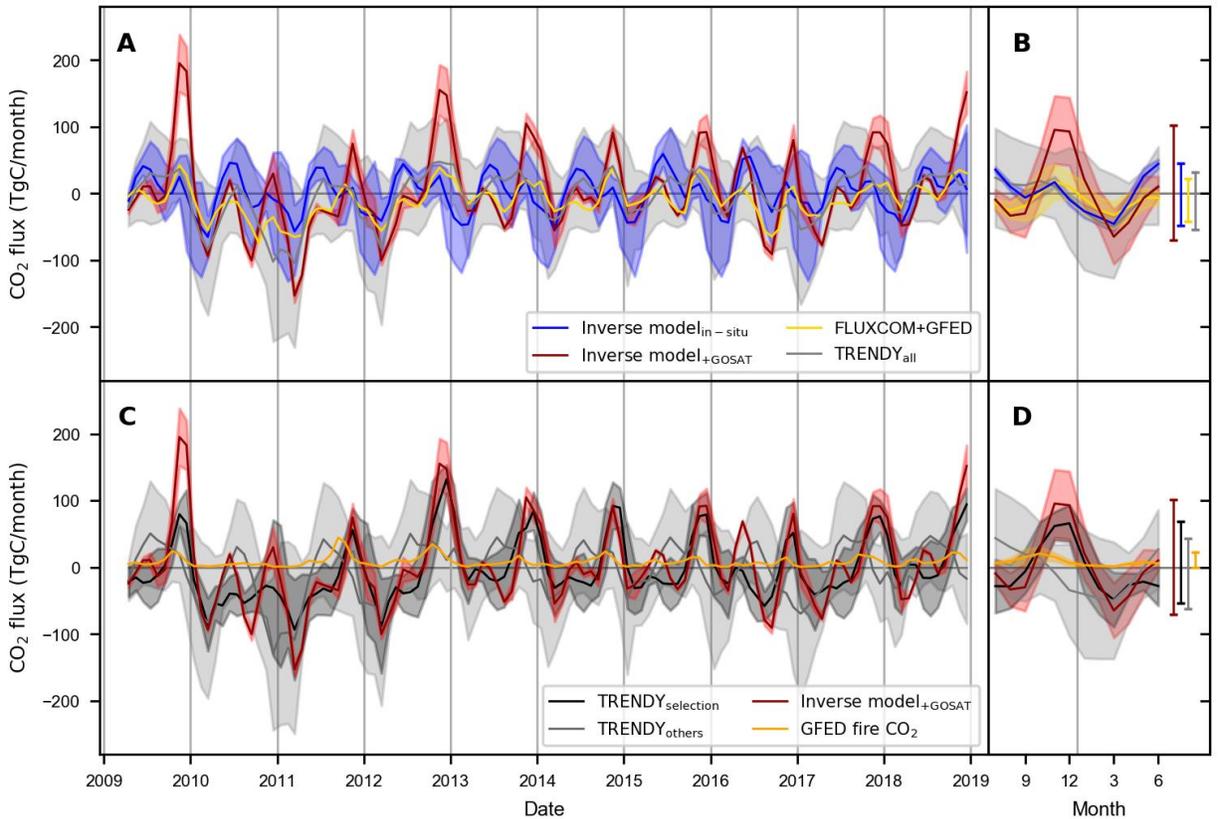

**Fig. 2. Australian net CO$_2$ fluxes. (A)** Top-down estimates of the net monthly Australian carbon fluxes inferred by in-situ CO$_2$ measurements based inverse models (blue) and by TM5-4DVAR assimilating in-situ measurements together with GOSAT observations (red), compared to bottom-up FLUXCOM+GFED NBP (yellow) and the TRENDY ensemble mean NBP (grey). Shading indicates the range among the various top-down data streams (in-situ based CarbonTracker, CAMS, and TM5-4DVAR in blue, TM5-4DVAR$_{+RemoTeC/GOSAT}$ and TM5-4DVAR$_{+ACOS/GOSAT}$ in red) and the standard deviation among the TRENDY ensemble (grey). **(C)** NBP of a subgroup of TRENDY models (black) compared to the other models (grey), to the GOSAT inversions (red, same as in (A)) and to GFED fire emissions (orange). Shading as in (A). **(B)** and **(D)** Mean and standard deviation (shading) over the period 2009 to 2018 and the mean peak-to-peak seasonal cycle amplitudes (bars). Positive fluxes correspond to carbon emissions into the atmosphere.

To understand the origin of the $CO_2$ pulses, we compare to bottom-up estimates from machine learning (FLUXCOM (*18, 20*)) and 18 process-based dynamic global vegetation models (DGVMs) from the TRENDY (v9) ensemble (*42*). Those also provide the component fluxes of gross primary productivity (GPP) and terrestrial ecosystem respiration (TER) enabling the attribution to variations in photosynthetic carbon uptake and respiratory carbon release. We further include fire emissions (FIRE) from the Global Fire Emission Database (GFED) as a potential factor for explaining the pattern. To compare to the top-down inversions, we calculate net biome production (NBP = TER + FIRE - GPP) by adding fire emissions from GFED to net ecosystem exchange (NEE = TER – GPP) from FLUXCOM. That is, positive fluxes correspond to carbon emissions into the atmosphere. For TRENDY, NBP is taken directly from the simulations of the DGVMs. We find that FLUXCOM+GFED derived NBP lacks the end-of-dry-season $CO_2$ pulses (Fig. 2A) and its seasonal amplitude (64±16 TgC/month) underestimates the one found by the GOSAT inversions by a factor of 3. This could be explained by the sparsity of Australian flux tower data in the training of the FLUXCOM machine learning models (only 4 of 224 sites lie in Australia, see Fig. S3) causing extrapolation errors (*18*), and by known weaknesses in representing certain fluctuations in response to water availability (*19*) or "memory" effects due to non-accounted carbon pool dynamics (*43*). Our analysis further suggests that local and transported fire emissions might contribute at the beginning of the carbon pulses but cannot explain their magnitude and duration (Fig. 2B and Fig. S5).

The ensemble of TRENDY NBP simulations shows a large inter-model spread and also no end-of-dry-season $CO_2$ pulses on average (Fig. 2A) causing a seasonal amplitude (85±20 TgC/month) which is about half of that of the GOSAT inversions. However, the dry season pulses are present in a subset of five of the TRENDY DGVMs (Fig. 2B and Table S1). For this subset, the timing and the magnitude (except for the year 2009) of the pulses and their seasonal amplitude (123±31 TgC/month) are closer to the pulses found by the GOSAT inversions. This finding suggests that the $CO_2$ pulses can be explained by ecosystem processes shaping the phasing of photosynthesis and respiration.

**Phasing of respiration and photosynthesis**

We find that the subset of DGVMs which are in good agreement with the GOSAT inversions reveals a distinctly different seasonal timing of GPP and TER than the other DGVMs. For the selected subset, the $CO_2$ pulses are driven by TER, which increases rapidly at the onset of the rainy season while GPP takes up only a few weeks later (Fig. 3A). The pulses originate from the semi-arid regions (Fig. S6). For the other DGMVs, TER and GPP show a mostly synchronous phasing throughout the year yielding no $CO_2$ pulses (Fig. 3B). The precipitation records for the semi-arid regions of Australia (Fig. 3C, Fig. S3) suggest that the respiration driven pulses shown by the GOSAT inversions and the selected TRENDY models are weaker or do not occur in years with anomalously strong precipitation during the dry period (Austral winter) such as in the La Nina years 2010 and 2016. This implies that the observed pulses are conditional on rewetting of dry soils.

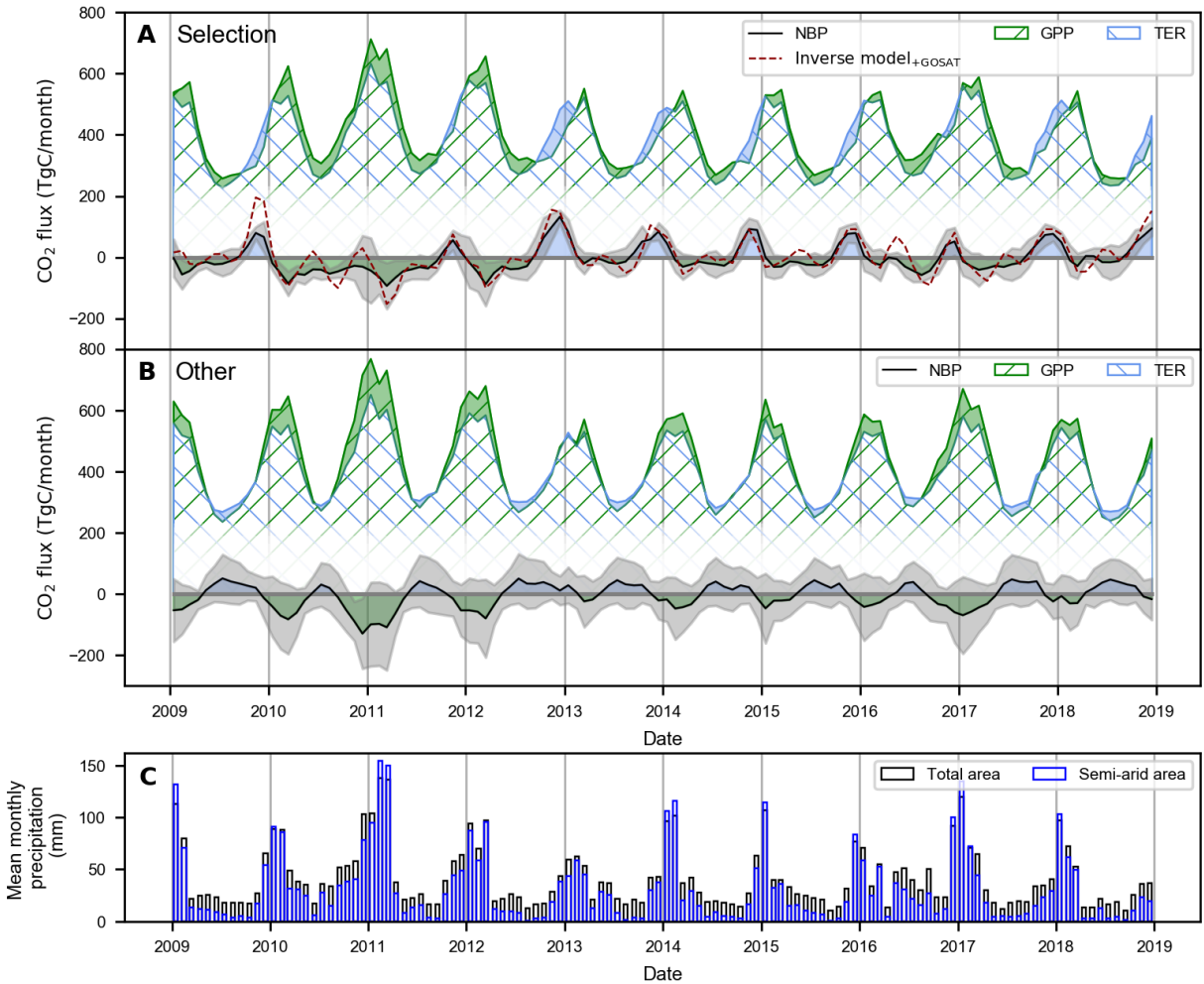

**Fig. 3. Seasonal timing of gross carbon fluxes among TRENDY models.** (**A**) Gross primary production (GPP, green) and total respiration (TER, light-blue) for Australia for the selection of TRENDY DGVMs that replicate the end-of-dry-season $CO_2$ pulses. NBP is shown in black in the lower part (grey shading indicates the standard deviation among the model subset). (**B**) Same as panel a but for the other TRENDY models that do not replicate the end-of-dry-season $CO_2$ pulses. (**C**) Mean monthly precipitation over the entire Australian region (black) and the semi-arid part (see Fig. S3) of Australia (blue).

With that, the detected continental-scale $CO_2$ pulses are consistent with site-level observations of dryland ecosystems which show an asynchronous response of respiration and photosynthesis to precipitation pulses (*44*). The rapid response of microbial respiration to rewetting events, is known as "Birch effect" and has been described in the literature of specific sites in some semi-arid regions for many decades (*29–31*). After being dormant in the dry period, soil microbes are activated by the moisture supply from rainfall. Benefitting from warm soils, accumulated and readily available substrate gets respired quickly going along with rapid growth of microbial populations. These dynamics of soil microbial processes cause a respiration $CO_2$ pulse with rewetting of dry soils which is also evident in Australian flux tower data (Fig. S7). Photodegradation of surface litter (*45*) and the death of microorganisms during the dry period (*46, 47*) may lead to the accumulation of easily decomposable substrate available to microorganisms at the onset of rain. It remains an open question whether the respiration pulses are mainly driven by substrates accumulated during the dry period and to what extent they are fueled by mobilization and decomposition of physically protected carbon (*47*). These processes are not represented explicitly or in detail in the TRENDY DGVMs and thus, the DGVMs cannot resolve how the site-level mechanisms scale up to the continental-scale effect observed here. Nonetheless, a selection of models effectively captures the continental-scale $CO_2$ pulses by a fast response of respiration and a delayed response of photosynthesis to the onset of the rainy season. This highlights the importance of subtle differences in effective parameterizations of respiration and photosynthesis to moisture fluctuations. Associated uncertainties affect the skill of the models to represent the carbon cycle of semi-arid ecosystems.

Our study demonstrates that the respiration driven $CO_2$ pulses over Australia following the end of the dry season are of large-scale relevance and appear to dominate the variability of the continent's carbon balance. It implies that GOSAT inversions have shed light on a blind spot of previous top-down and bottom-up approaches for quantifying and attributing $CO_2$ flux variability over semi-arid regions. This calls for revisiting the contribution of semi-arid systems to $CO_2$ balance variations on global scales. Considering changing precipitation patterns under climate change, the suggested continental-scale process understanding may improve representations of climate-carbon cycle feedbacks and projections of future carbon fluxes.

**Materials and Methods**

Summary of observation and model data

The main characteristics of the observation and model data are listed in Table S1.

TRANSCOM region Australia

Our region of interest is 'Australia' as defined by the TRANSCOM-3 experiment (*48*) including the Australian continent and New Zealand. For the main analysis, concentration and flux data are averaged and aggregated, respectively, over a month or a year for the entire region. Satellite concentrations are only reported if averaging includes more than 10 data points. To avoid sampling effects on the coastline, all flux datasets are aggregated on a 1°×1° grid before applying the TRANSCOM region mask to aggregate over the entire region and one month. Grid cells with their centers inside the Australian region are counted to belong to the region.

$CO_2$ concentrations

We primarily use GOSAT column-average dry-air mole fractions of $CO_2$ (Fig. 1), also denoted $XCO_2$, generated by operating the RemoTeC radiative transfer and retrieval algorithm (*8, 34*) on shortwave-infrared spectra of sunlight backscattered to GOSAT by the Earth's surface and atmosphere (called GOSAT/RemoTeC). The algorithm version employed here corresponds to the one used previously (*8*) with updates related to the quality filtering and to ancillary input data, in particular updated a priori gas concentrations. Furthermore, we also use GOSAT $CO_2$ records generated by the NASA Atmospheric $CO_2$ Observations from Space (ACOS) algorithm version 9r(Lite) (*49*) (called GOSAT/ACOS).

To confirm robustness of the satellite data, we compare GOSAT $CO_2$ against records of the Orbiting Carbon Observatory-2 version 10 (OCO-2) (*50*) covering the time period 2014 to 2018 (Table S1 and Fig. 1). We further compare the satellite data to ground-based measurements of the column-average dry-air mole fractions reported by the Total Carbon Column Observing Network (TCCON) (*38*). Thereby, data of the two Australian stations Darwin and Wollongong are used (Table S1 and Fig. S2). Both stations are located near the coastline and neither are in the semi-arid regions (see Fig. S3). Therefore, the comparison to the continental GOSAT data suffers from limited representativeness.

Simulated $CO_2$ concentrations (Fig. 1) are taken from three inverse atmospheric transport models (Table S1) that estimate surface-atmosphere fluxes which are optimally compatible with atmospheric concentration measurements and prior flux knowledge: TM5 four-dimensional variational inversion system (TM5-4DVAR) (*41*), CarbonTracker (CT2019B) (*39, 51*), and the Copernicus Atmosphere Monitoring Service (CAMS) (*40, 52, 53*). Given the optimized fluxes, the transport model is run forward to produce simulated concentration fields. All three models assimilate ground-based in-situ $CO_2$ concentration measurements collected from the global monitoring networks (*54*). We use TM5-4DVAR for further analysis to assimilate the GOSAT/RemoTeC, GOSAT/ACOS, and OCO-2 $CO_2$ data together with the in-situ observations.

For illustrating the seasonal concentration dynamics in Fig. 1, we remove the secular increase of $CO_2$ concentrations in the atmosphere by detrending the concentration data, i.e. we subtract the global atmospheric background assuming a piece-wise yearly linear increase according to the annual mean carbon dioxide growth rates (*GR*) reported by the National Oceanic and Atmospheric Administration (NOAA) based on globally averaged marine surface data (*55*). Thus, the background concentration for month *m* ([1,...,12]) and year *y* ([2009,...,2018]) reads:

$$BG_{y,m} = BG + \sum_{i=2009}^{y-1}(GR_i) + \frac{m}{12}GR_y \qquad (1)$$

where *BG* is an overall offset determined by setting the mean of the detrended $CO_2$ concentrations to zero, the second term accumulates the growth since the start of the time series in the year 2009 until the start of year *y*, and the third term accounts for the fractional increase during the respective year *y*. We subtract the background individually for all $CO_2$ concentration data sets (satellite as well as simulation data). Note that detrending is only applied to concentration data used in Fig. 1 for illustration purposes, the inverse models assimilate whole $CO_2$ concentrations.

$CO_2$ top-down fluxes

The three inverse atmospheric models TM5-4DVAR, CarbonTracker, and CAMS, that provide simulated $CO_2$ concentration fields, also provide estimates of the surface-atmosphere fluxes compatible with ground-based in-situ $CO_2$ measurements (Fig. 2A). For further analysis, we use TM5-4DVAR to assimilate the GOSAT $CO_2$ data together with the ground-based in-situ observations (Fig. 2A and 2B). Furthermore, we assimilate OCO-2 data together with in-situ measurements to obtain fluxes for comparison (Fig. S4). Depending on whether GOSAT/RemoTeC, GOSAT/ACOS, or OCO-2 data are used, we denote the respective flux estimates in the Extended Materials with InverseModel$_{+RemoTeC/GOSAT}$, InverseModel$_{+ACOS/GOSAT}$, and InverseModel$_{+OCO-2}$. The models provide output in terms of the net $CO_2$ fluxes partitioned into biosphere, oceanic, fire, and fossil fluxes. TM5-4DVAR is configured to estimate weekly biosphere and oceanic fluxes on a regular 6°(longitude) × 4°(latitude) grid while fire and fossil emissions are imposed from the Quick Fire Emissions Dataset (QFED (*56*)) and the Open-source Data Inventory for Anthropogenic $CO_2$ (ODIAC (*57, 58*)), respectively. The construction of the prior oceanic, fire and biosphere fluxes are detailed elsewhere (*59*). We average the oceanic, biospheric and fire fluxes between 2000 and 2019 to create 20-year climatological land and ocean sinks. We then apply year-specific scaling on these sinks to match the observed annual atmospheric $CO_2$ growth given year-specific fossil $CO_2$ emissions. The prior fluxes thus constructed follow the atmospheric growth of $CO_2$ over two decades.

For all inversions, NBP is calculated as the sum of a posteriori biosphere fluxes and fire emissions. Positive fluxes correspond to carbon emissions into the atmosphere, negative fluxes indicate carbon uptake by the ecosystems. While all TM5-4DVAR data is already provided on the scale of TRANSCOM regions, CAMS and CarbonTracker fluxes are aggregated on a 1°x1° grid before applying the TRANSCOM region mask.

$CO_2$ bottom-up fluxes

FLUXCOM provides estimates of global bottom-up net ecosystem exchange (NEE) based on upscaling of local flux measurements. To this end, a machine learning approach uses the eddy covariance measurements by the FLUXNET tower network together with meteorological and satellite remote sensing data to deliver NEE globally at fine spatial resolution (*18, 20*). The FLUXCOM version, used here, only includes four stations of the Australian OzFlux network (Fig. S3). To calculate FLUXCOM compatible NBP (Fig. 2A), we take the sum of the remote sensing FLUXCOM ensemble and fire emissions from the Global Fire Emission Database (GFED) v4.1s (*60*). Fluxes due to land-use change are neglected.

The TRENDY model inter-comparison project collects various DGVMs and contributes to the Global Carbon Project (*1*). Here, we use 18 TRENDY version 9 models listed in Table S1. NBP, GPP and TER provided by the TRENDY DGVMs are aggregated on a 1°x1° grid before applying the TRANSCOM region mask. As the land-ocean masks among the TRENDY models differ, the continental NBP is taken as mean flux in units $\mu g CO_2 m^{-2} s^{-1}$, then multiplied by the Australian region area to obtain total fluxes and converted to TgC/month. Most of the models provide NBP directly. For the models CABLE-POP and DLEM, not providing net fluxes, NBP is constructed from only GPP and TER, as both models do not provide FIRE fluxes. The subset of models showing the end-of-dry-season $CO_2$ pulses is termed TRENDY$_{selection}$. The other subset of TRENDY models not showing the pulses are called TRENDY$_{others}$ (Fig. 2B, Fig. 3 and Table S1).

Figure 3C shows the timing of bottom-up NBP for correlations with monthly mean precipitation. The latter is taken from the European Centre for Medium Range Weather Forecasts (ECMWF) ERA5-land data product (*61, 62*). We average the ERA-5 data over entire Australia and the semi-arid parts (see Fig. S3) defined as all the 1°x1° grid cells with less than 22 mm of monthly mean precipitation during four consecutive months in the ten-year averaged annual cycle.

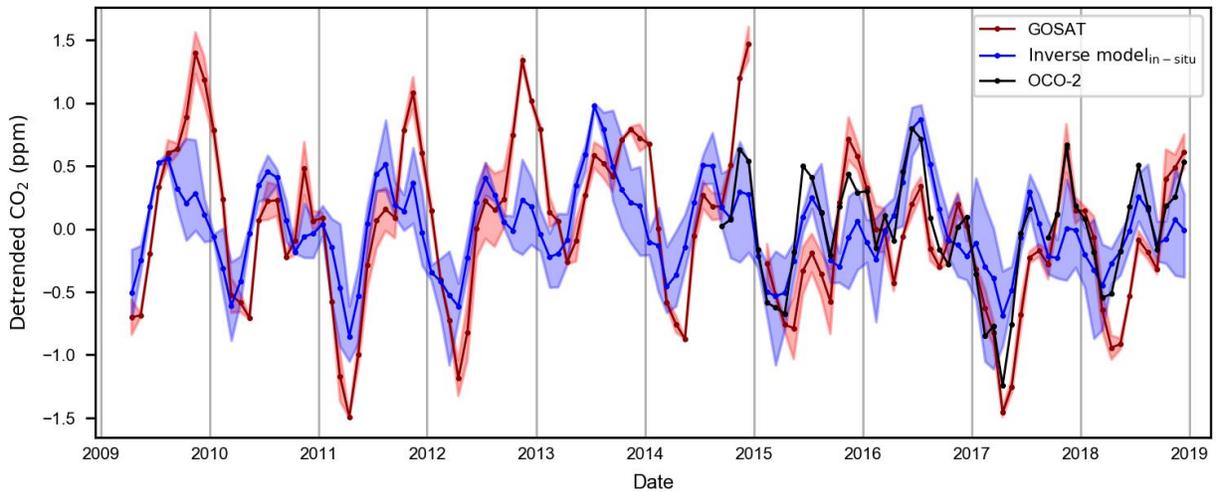

**Fig. S1. Detrended $CO_2$ concentrations above Australia from GOSAT, OCO-2 and inverse models.** Detrended monthly mean column-average dry-air mole fractions of $CO_2$ measured by GOSAT (red), OCO-2 (black, from 2014) and simulated by in-situ-driven inverse models (blue) averaged over continental Australia. Red shading indicates the range of the GOSAT/RemoTeC and GOSAT/ACOS algorithms. Blue shading indicates the range of the CarbonTracker, CAMS, and TM5-4DVAR inverse models.

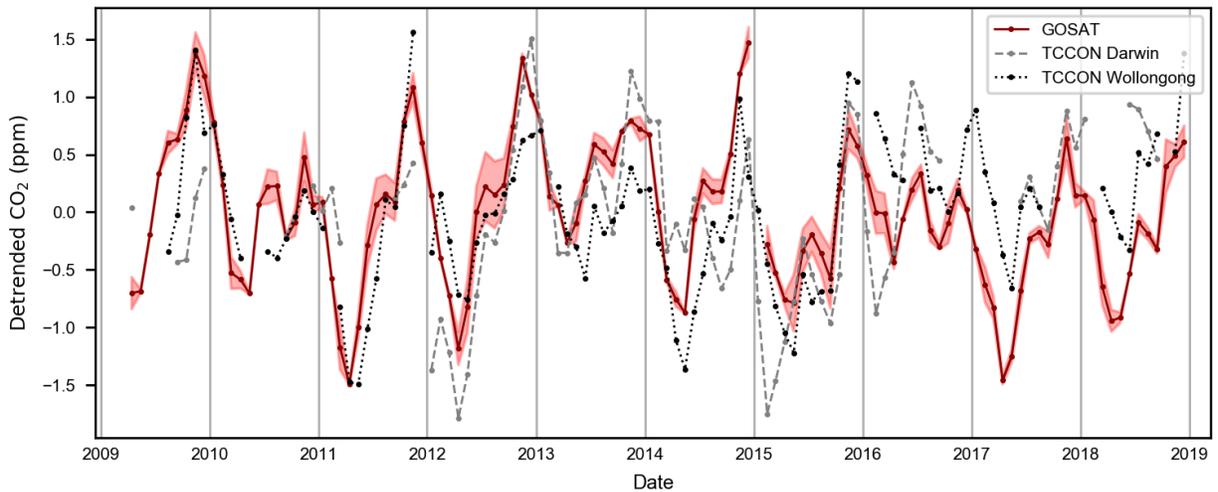

**Fig. S2. Detrended $CO_2$ concentrations above Australia from satellite and TCCON stations.** Detrended monthly mean column-average dry-air mole fractions of $CO_2$ measured by GOSAT (red) averaged over continental Australia and for individual TCCON stations (Darwin (*63*) in grey, Wollongong (*64*) in black). Red shading indicates the range of the GOSAT/RemoTeC and GOSAT/ACOS algorithms.

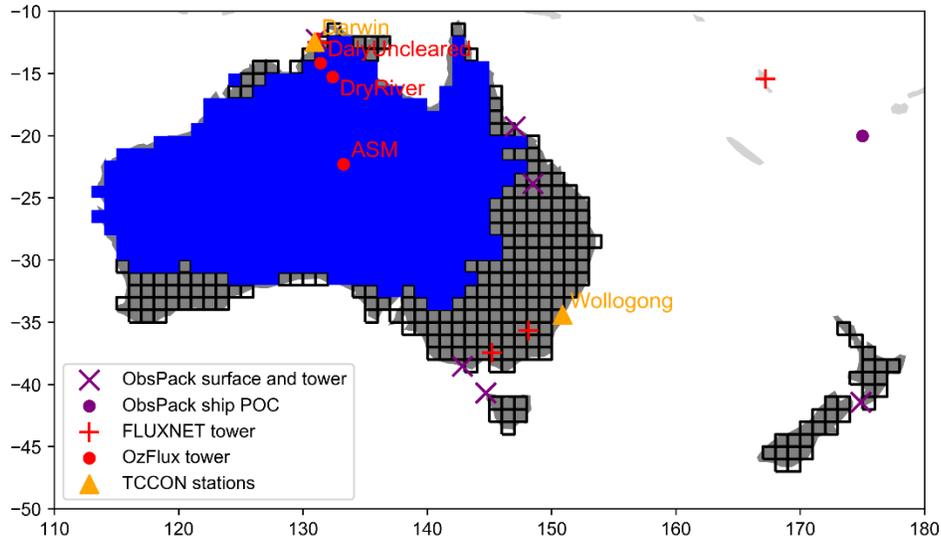

**Fig. S3. TRANSCOM region and $CO_2$ measurement stations.** The Australian regions of the TRANSCOM-3 intercomparison project is depicted in dark grey. The TRANSCOM region Australia includes Australia and New-Zealand and is divided in a semi-arid (blue) and not semi-arid part (black borders) on a 1°x1° grid. The $CO_2$ concentration measurement stations included in ObsPack (*54*) are shown in purple (crosses for surface and tower measurements, dot for Pacific Ocean Cruise (POC) measurements). These measurements are used by the inverse models. The eddy covariance flux measurement towers within FLUXNET and used by FLUXCOM are given as red crosses. The three OzFlux towers used in Fig. S6 are given as red dots with labels. The two TCCON stations are marked as yellow triangles with labels.

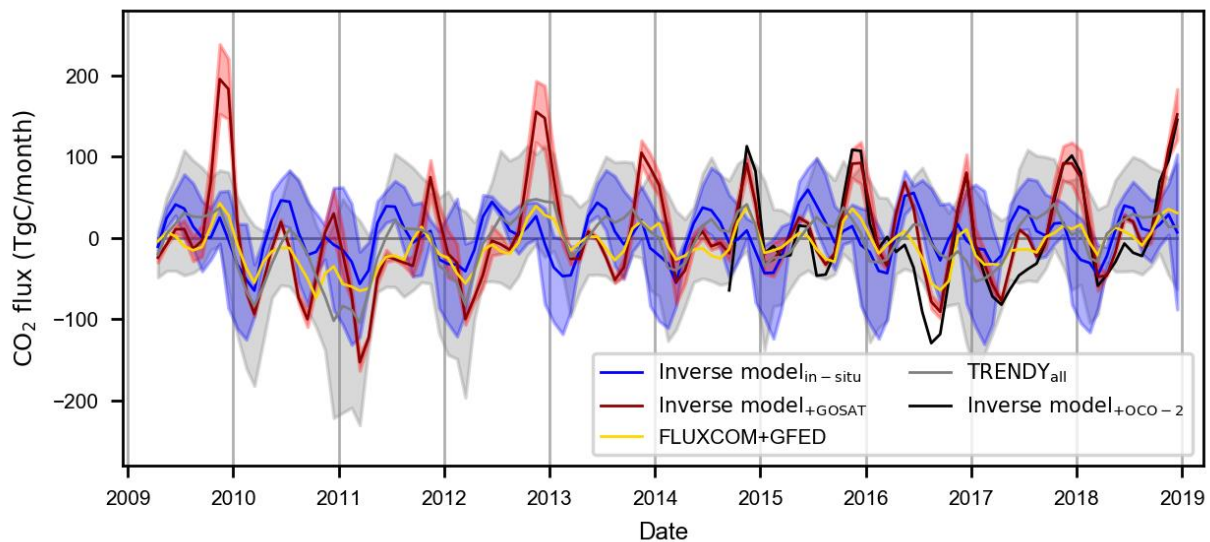

**Fig. S4. Australian net $CO_2$ fluxes with OCO-2 based fluxes.** Like Fig. 2A, but additionally with OCO-2 based fluxes. Top-down estimates of the net monthly Australian carbon fluxes inferred by TM5-4DV AR assimilating in-situ $CO_2$ measurements alone (blue), together with GOSAT observations (red), and together with OCO-2 (black), compared to bottom-up FLUXCOM+GFED NBP (yellow) and the TRENDY ensemble mean NBP (grey). Shading indicates the range among the various top-down data streams (blue, red) and the standard deviation among the TRENDY ensemble (grey).

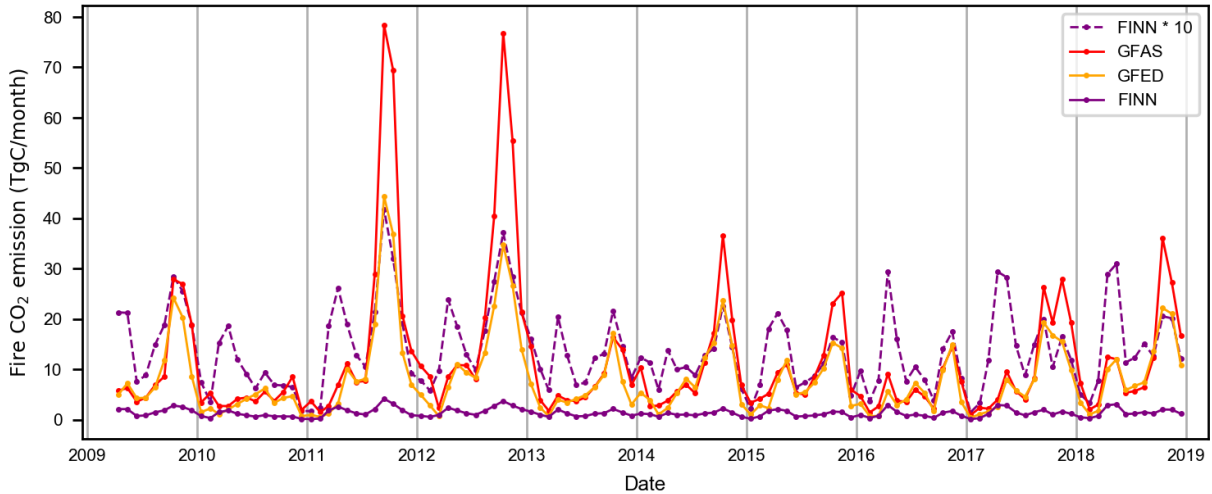

**Fig. S5. CO$_2$ fire emissions in Australia.** The monthly CO$_2$ fire emissions collected by three fire emission databases (GFED in orange, Global Fire Assimilation System (GFAS (*65*)) in red and the Fire INventory from NCAR (FINN (*66*)) in purple). The FINN fire emissions are additionally given amplified by a factor of ten to visualize their seasonal structure.

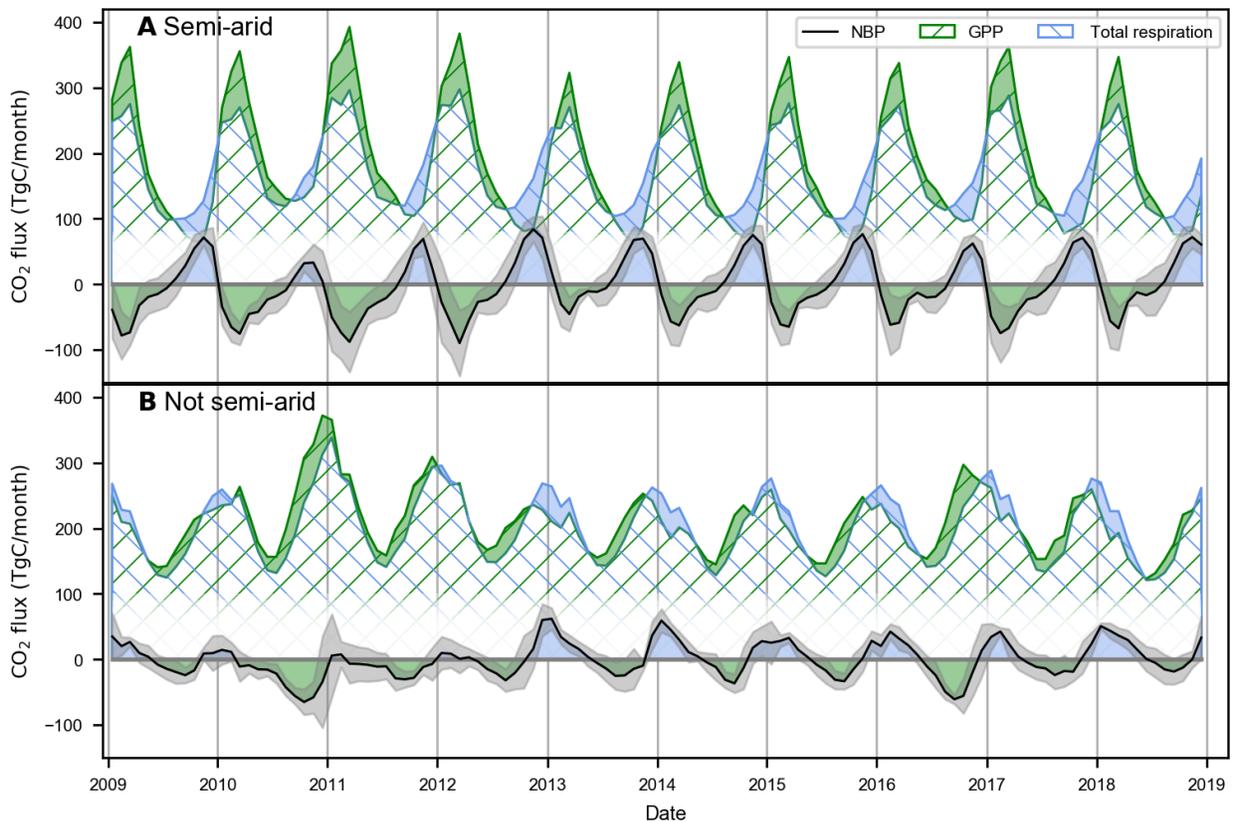

**Fig. S6. Seasonal timing of gross carbon fluxes among the selected TRENDY models**. **(A),** Gross primary production (GPP, green) and total respiration (TER, light-blue) for the semi-arid parts of Australia (see map Fig. S3) for the selection of TRENDY DGVMs that replicate the end-of-dry-season CO$_2$ pulses. NBP is shown in black in the lower part (grey shading indicates the standard deviation among the model subset). **(B),** Same as panel (A) but for the parts of Australia which are not semi-arid.

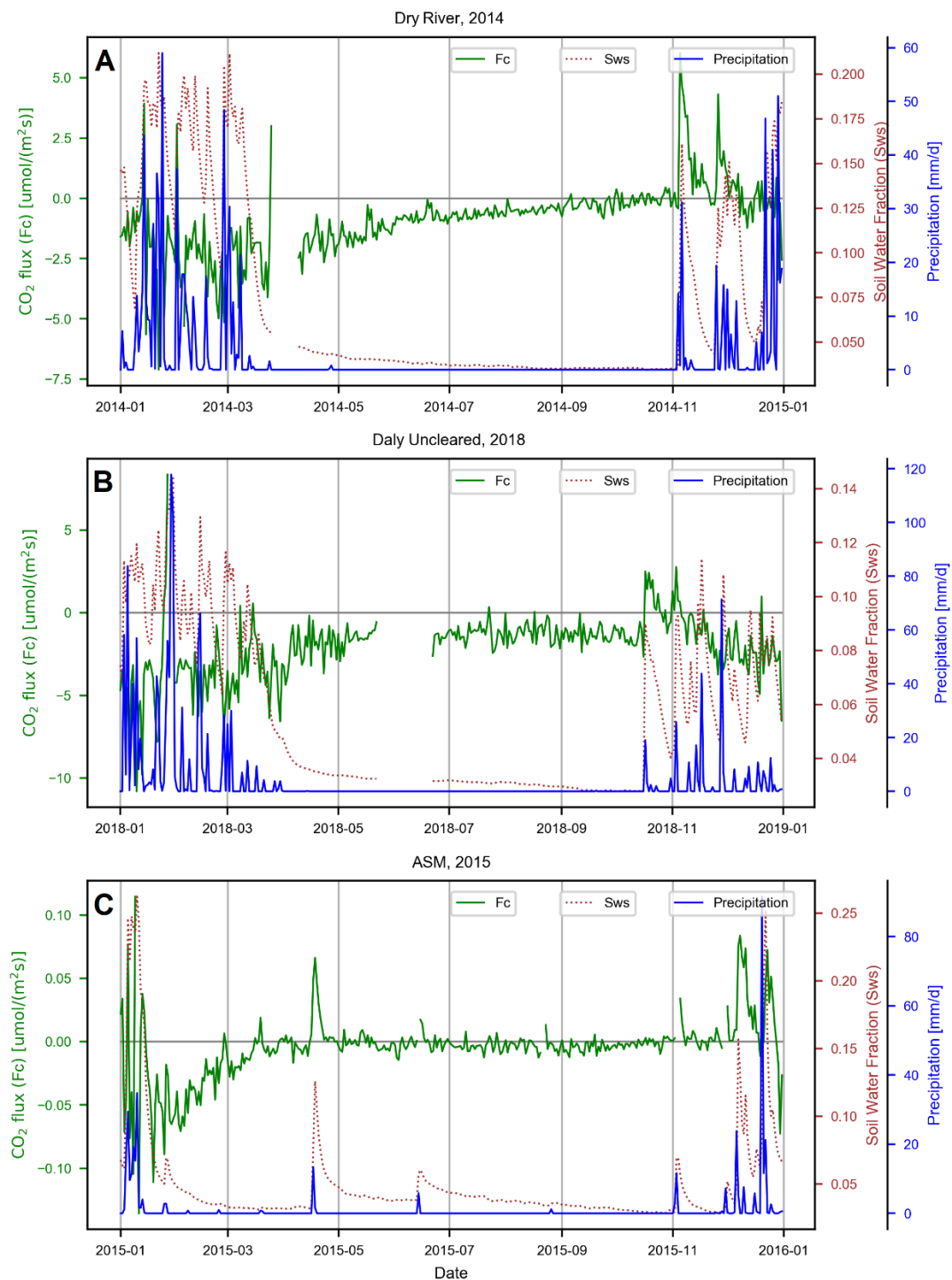

**Fig. S7. Local data from OzFlux eddy covariance flux towers. (A)-(C)** Daily mean net carbon fluxes (FC in green), precipitation (blue) and soil moisture (red dashed) measured by OzFlux stations for periods illustrating local correlations between moisture supply and $CO_2$ fluxes. (A) Station record Daly Uncleared (*67*). (B) Station record Dry River (*68*). (C) Station record Alice Springs Mulga (*69*) (ASM). The locations of the stations are given in Fig. S3.

**Table S1.** Summary of datasets.

| Description | Dataset | Resolution | References |
|---|---|---|---|
| GOSAT XCO$_2$ | GOSAT/RemoTeC v2.4.0<br>GOSAT/ACOS v9r(Lite) | 10.5 km footprint<br>10.5 km footprint | (34)<br>(70) |
| Validation XCO$_2$ | OCO-2 v10r<br>TCCON Darwin, Wollongon | 1.3×2.3 km footprint<br>local | (50, 71)<br>(38, 63, 64) |
| Model XCO$_2$<br>based on in-situ data | TM5 − 4DVAR$_{in-situ}$<br>CarbonTracker CT2019B$_{in-situ}$<br>CAMS$_{in-situ}$ v20r2 | regional, monthly<br>3°×2°, monthly<br>3.7°×1.81°, monthly | (41)<br>(51)<br>(40, 52, 53) |
| Inverse Model$_{in-situ}$ | TM5 − 4DVAR$_{in-situ}$<br>CarbonTracker CT2019B$_{in-situ}$<br>CAMS$_{in-situ}$ v20r2 | regional, monthly<br>1°×1°, monthly<br>3.7°×1.81°, monthly | (41)<br>(51)<br>(40, 52, 53) |
| Inverse Model$_{+GOSAT}$ | TM5-4DVAR/RemoTeC<br>TM5-4DVAR/ACOS | regional, monthly<br>regional, monthly | (41)<br>(41) |
| FLUXCOM<br>+ GFED | FLUXCOM NEE<br>GFED v4.1s | 0.08°×0.08°, 8-days<br>0.25°×0.25°, monthly | (18, 20)<br>(72) |
| TRENDY$_{selection}$ | JSBACH S3<br>CLASSIC S3<br>LPJ S3<br>YIBs S3<br>OCN S3 | 1.86°x1.88° [1)]<br>2.80°x2.81° [1)]<br>0.5°x0.5° [1)]<br>1°x1° [1)]<br>1°x1° [1)] | (73)<br>(74)<br>(75)<br>(76)<br>(77) |
| TRENDY$_{others}$ | ORCHIDEE-CNP S3<br>ORCHIDEE S3<br>ORCHIDEEv3 S3<br>CABLE-POP S3<br>CLM5.0 S3<br>DLEM S3<br>IBIS S3<br>ISAM S3<br>ISBA-CTRIP S3<br>JULES-ES-1.0 S3<br>LPX-Bern S3<br>SDGVM S3<br>VISIT S3 | 2°x2° [1)]<br>0.5°x0.5° [1)]<br>2°x2° [1)]<br>1°x1° [1)]<br>0.94°x1.25° [1)]<br>0.5°x0.5° [1)]<br>1°x1° [1)]<br>0.5°x0.5° [1)]<br>1°x1° [1)]<br>1.25°x1.88° [1)]<br>0.5°x0.5° [1)]<br>1°x1° [1)]<br>0.5°x0.5° [1)] | (78)<br>(79)<br>(80)<br>(81)<br>(82)<br>(83)<br>(84)<br>(85)<br>(86)<br>(87)<br>(88)<br>(89)<br>(90) |
| precipitation | ERA5-land data<br>total precipitation | 1°×1°, monthly | (61, 62) |

[1)] all TRENDY model data is provided in monthly temporal resolution
The main characteristics and references of the observation and model data are listed. Links to the data-sets are provided in the 'Availability of data and materials' section.

**Table S2.** Seasonal and interannual variability of CO$_2$ flux datasets.

| Ensembles | Mean Amplitude [TgC/month] | Relative Amplitude | Standard Deviation [TgC/month] | IAV [TgC/a] |
|---|---|---|---|---|
| Inv. Model$_{+GOSAT}$ | 172.00 | 1 | 46.52 | 233.26 |
| Inv. Model$_{in-situ}$ | 93.17 | 0.54 | 11.05 | 48.04 |
| TRENDY$_{all}$ | 85.40 | 0.50 | 20.09 | 210.17 |
| TRENDY$_{selection}$ | 122.95 | 0.71 | 30.51 | 236.03 |
| TRENDY$_{others}$ | 104.83 | 0.61 | 27.00 | 201.15 |
| FLUXCOM+GFED | 64.09 | 0.37 | 15.85 | 157.45 |
| GFED | 21.82 | 0.13 | 10.16 | |

July-to-June peak-to-peak amplitude of NBP (mean in TgC/month, relative w.r.t. the GOSAT inversions, standard deviation in TgC/month over the 2009 to 2018 period) and NBP interannual variations (IAV) (standard deviation in TgC/a over the 2009 to 2018 period) for the datasets used.

89. A. P. Walker, T. Quaife, P. M. van Bodegom, M. G. de Kauwe, T. F. Keenan, J. Joiner, M. R. Lomas, N. MacBean, C. Xu, X. Yang, F. I. Woodward, The impact of alternative trait-scaling hypotheses for the maximum photosynthetic carboxylation rate (Vcmax) on global gross primary production. *The New phytologist*. **215**, 1370–1386 (2017), doi:10.1111/nph.14623.
90. E. Kato, T. Kinoshita, A. Ito, M. Kawamiya, Y. Yamagata, Evaluation of spatially explicit emission scenario of land-use change and biomass burning using a process-based biogeochemical model. *Journal of Land Use Science*. **8**, 104–122 (2013), doi:10.1080/1747423x.2011.628705.
**Acknowledgments:** We thank the Japanese Aerospace Exploration Agency, National Institute for Environmental Studies and the Ministry of Environment for the GOSAT data and their continuous support as part of the Joint Research Agreement. OCO-2 data were produced by the OCO-2 project at the Jet Propulsion Laboratory, California Institute of Technology, and obtained from the OCO-2 data archive maintained at the NASA Goddard Earth Science Data and Information Services Center. CarbonTracker CT2019B results are provided by NOAA ESRL, Boulder, Colorado, USA from the website at http://carbontracker.noaa.gov. We thank all TRENDY modelers for providing model output as part of the TRENDY v9 ensemble. The study has greatly benefited from discussions with Christian Frankenberg.**Author contributions:**

AB, SNV, and EMM were involved in conceptualization and methodology. EMM conducted the formal analysis and the visualization under supervision of AB and SNV. AB, SNV, EMM, MJ, and SB wrote the original draft. SB performed the dedicated TM5-4DVar runs. SS, VKA, PRB, PF, DSG, AKJ, EK, JEMSN, BP, RS, HT, AW, WY, XY, SZ provided TRENDY data. NMD and DWTG provided TCCON data. All authors contributed to the editing and review of the manuscript.

**Competing interests:** Authors declare that they have no competing interests.

**Data and materials availability:** GOSAT/RemoTeC2.4.0 XCO2 data can be obtained from doi: 10.5281/zenodo.5886662 (last access: 2022-02-25). GOSAT/ACOS data is available at https://oco2.gesdisc.eosdis.nasa.gov/data/GOSAT_TANSO_Level2/ACOS_L2_Lite_FP.9r/ (last access: 2020-07-28). OCO-2 data is available at https://disc.gsfc.nasa.gov/datasets/OCO2_L2_Lite_FP_10r/summary (last access: 2020-11-01). TCCON data can be downloaded at https://data.caltech.edu/records/269 (last access: 2022-02-25). CarbonTracker CT2019B CO2 fluxes and concentrations can be downloaded from https://gml.noaa.gov/aftp/products/carbontracker/co2/CT2019B/fluxes/monthly/ (last access: 2021-02-19) and https://gml.noaa.gov/aftp/products/carbontracker/co2/CT2019B/molefractions/co2_total_monthly/ (last access: 2022-02.25) respectively. CAMS concentrations and fluxes can be found at datasets/data/cams-ghg-inversions/ (last access: 2021-10-07). GFAS emissions records are available at https://apps.ecmwf.int/datasets/data/cams-gfas/ (last access: 2020-11-13). CAMS and GFAS data were generated using Copernicus Atmosphere Service Information [2021] and neither the European Commission nor ECMWF is responsible for any use that may be made of the information it contains. GFED fire emissions are available at https://www.geo.vu.nl/~gwerf/GFED/GFED4/ (last access: 2020-07-10). FINN data were retrieved from the American National Center for Atmospheric Research https://www2.acom.ucar.edu/modeling/finn-fire-inventory-ncar (last access: 2020-11-18). The used OzFlux data can be downloaded from https://www.ozflux.org.au/ (last access: 2021-11-16). ERA5-land data records contain modified Copernicus Atmosphere Service Information [2021] available at the Climate Data Store https://cds.climate.copernicus.eu/cdsapp#!/dataset/reanalysis-era5-land-monthly-means (last access: 2021-12-20). TRENDYv9 model output and FLUXCOM products are available upon request (https://sites.exeter.ac.uk/trendy and http://fluxcom.org/CF-Download/ respectively). The TM5-4DVar data can be requested from the corresponding authors.

The code used in this study is available from the corresponding authors on reasonable request.